\documentclass[prl,aps,reprint,showpacs,superscriptaddress,amsmath]{revtex4-1}

\usepackage{bm}
\usepackage{graphicx}
\usepackage{dsfont}

\newcommand{\bra}[1]{\langle #1 | \,}
\newcommand{\ket}[1]{\, | #1 \rangle}

\newcommand{\expv}[1]{\langle #1 \rangle}
\newcommand{\bs}[1]{\boldsymbol{#1}}
\newcommand{\lra}{\leftrightarrow}
\newcommand{\om}{\omega}
\newcommand{\Om}{\Omega}
\newcommand{\ga}{\gamma}
\newcommand{\Ga}{\Gamma}
\newcommand{\de}{\delta}
\newcommand{\De}{\Delta}

\newcommand{\la}{\lambda}
\newcommand{\eps}{\epsilon}
\newcommand{\veps}{\varepsilon}
\newcommand{\hlf}{\frac{1}{2}}
\newcommand{\mc}[1]{\mathcal{#1}}
\newcommand{\sig}{\hat{\sigma}}

\newcommand{\hrho}{\hat{\rho}}
\newcommand{\hL}{\hat{L}}

\begin{document}

\title{Binding Potentials and Interaction Gates 
between Microwave-Dressed Rydberg Atoms}

\author{David Petrosyan}
\affiliation{Aarhus Institute of Advanced Studies, Aarhus University,
DK-8000 Aarhus C, Denmark}
\affiliation{Institute of Electronic Structure and Laser, FORTH,
GR-71110 Heraklion, Crete, Greece}

\author{Klaus M\o lmer}
\affiliation{Department of Physics and Astronomy, Aarhus University,
DK-8000 Aarhus C, Denmark}

\date{\today}

\begin{abstract}
We demonstrate finite range binding potentials between pairs
of Rydberg atoms interacting with each other via attractive and
repulsive van der Waals potentials and driven by a microwave field.
We show that, using destructive quantum interference to cancel single-atom
Rydberg excitation, the Rydberg-dimer states can be selectively and
coherently populated from the two-atom ground state. This can be used
to realize a two-qubit interaction gate which is not susceptible to mechanical
forces between the atoms and is therefore immune to motional decoherence.
\end{abstract}


\maketitle

Atoms in high-lying Rydberg states exhibit many remarkable features,
including long lifetimes and giant polarizability \cite{RydAtoms}.
The resulting strong, long-range, resonant (F\"orster) and nonresonant
(van der Waals) dipole-dipole interactions between the atoms can
suppress multiple Rydberg excitations within a certain blockade
distance \cite{Jaksch2000,Lukin2001,rydQIrev,rydDBrev}. 
In combination with laser and microwave field manipulation of atomic 
states, these interactions form the basis for quantum information 
processing \cite{rydQIrev} with individual atoms
\cite{Jaksch2000,NatPRLSaffman,NatPRLGrangier,Beguin2013,Durga2014} 
and atomic ensembles \cite{Lukin2001,rydQIrev,Kuzmich1213}.
Furthermore, cold atoms excited to Rydberg states represent a flexible 
platform to simulate \cite{RydQSim2010} and study few-body
\cite{Greene2000,RyGrmolexp,Boisseau2002,Overstreet2009,Samboy2011,%
Kiffner2012,Kiffner2013,Yu2013,AtesOlmos2012,WLi2013} and many-body physics
\cite{Pupillo2010,Weimer2008,Weimer10,Pohl2010,Schachenmayer2010,%
Bijnen2011,Lesanovsky2011,Lesanovsky2012,Garttner2012,Hoening2013,Zeller2012,
Ji2011,Ates2012,DPMHMF2013,Petrosyan2013,Lesanovsky2013,Viteau2011,Schauss2012}.

A paradigmatic interaction phenomenon is the formation of a bound pair 
of particles. An electron in a Rydberg orbit scattering off a ground-state
atom can sustain a weakly bound molecule with permanent dipole 
moment \cite{Greene2000,RyGrmolexp}. Macrodimers of two Rydberg atoms
can form due to van der Waals (vdW) interactions in a static electric 
field \cite{Boisseau2002,Overstreet2009} or through mixing of orbital 
angular momentum states \cite{Samboy2011}. Another binding mechanism 
relies on the Stark-shifted dipole-dipole (DD) interactions which can 
support Rydberg dimers \cite{Kiffner2012} and trimers \cite{Kiffner2013}.

Here we identify a new mechanism to obtain long-range binding potentials
between Rydberg atoms. Our two-atom potential curves result from microwave
field coupling between a pair of Rydberg states of each atom \cite{Yu2013}.
Atoms in these states interact with each other via repulsive and attractive
vdW potentials, and via typically weaker DD exchange interaction on the 
allowed microwave transition. When the attractive vdW interaction is 
comparable to, or stronger than, the repulsive vdW (and DD) interaction,
the microwave-dressed potential energy curves have pronounced wells
located in the vicinity of crossings of the two-atom attractive and 
repulsive potentials, in the frame rotating with the microwave frequency.
The microwave field, which is detuned from the transition resonance of 
a single atom, lifts the degeneracy and causes level anti-crossing in 
the two atom basis. We note that the combination of resonant and 
non-resonant DD interactions between the Rydberg atoms can also result 
in a binding potential \cite{Kiffner2012}.

We next address the question of how to selectively populate these
Rydberg-dimer states starting from the two-atom ground state.
The strong microwave field induces a broad dark resonance \cite{stirap}
for the laser excitation of a single (non-interacting) atom. We show
that within this electromagnetically induced transparency (EIT) \cite{EIT}
window, a smooth probe laser pulse populates only the two-atom 
Rydberg manifold. We then propose to employ such a Rydberg dimer state 
to realize the universal \textsc{cphase} quantum logic gate between a 
pair of qubits represented by atoms trapped at a suitable distance 
from each other. This is implemented by coherent excitation and de-excitation
of the atoms with a probe pulse of effective area $2\pi$. The quantum 
interference responsible for EIT prevents Rydberg excitation of a single atom,
and only an appropriate two-atom state acquires a conditional phase $\pi$.
Since during the gate operation we populate the two-atom Rydberg state
at the bottom of a potential well, there is no mechanical force between
the atoms and motional decoherence \cite{WLi2013} is suppressed.

\begin{figure}[t]
\centerline{\includegraphics[width=8.7cm]{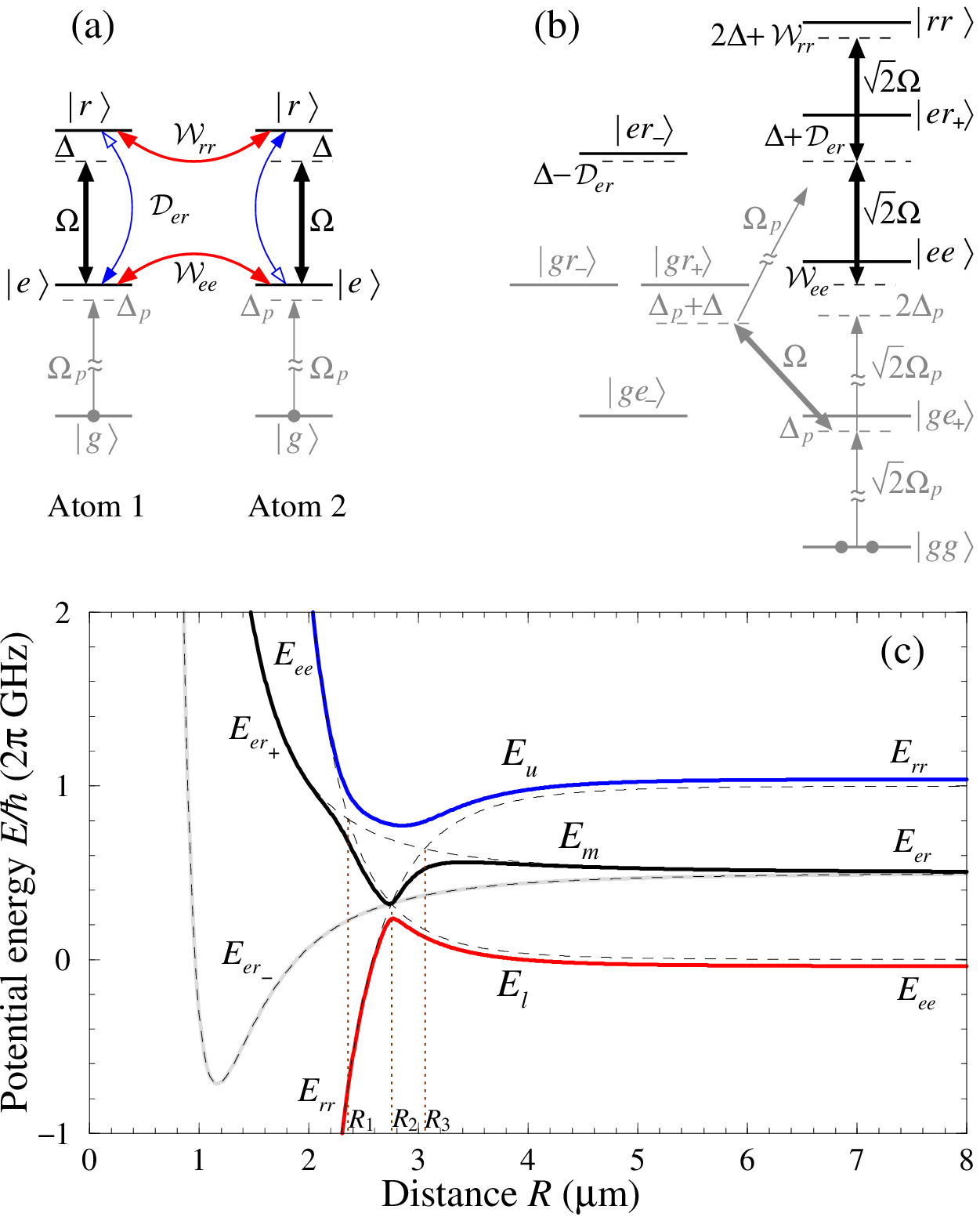}}
\caption{
(a) Atoms 1 and 2 with Rydberg states $\ket{e}$ and $\ket{r}$ interact
via repulsive $\mc{W}_{ee}$ and attractive $\mc{W}_{rr}$ vdW interactions
and DD (exchange) interaction $\mc{D}_{er}$, while a microwave field 
drives transition $\ket{e} \lra \ket{r}$ with Rabi frequency
$\Om$ and detuning $\De$. The Rydberg states can be excited
from the ground state $\ket{g}$ by a probe laser field with
Rabi frequency $\Om_p$ and detuning $\De_p$.
(b) Diagram of levels and couplings in the two-atom basis. 
Antisymmetric states  $\ket{ge_-}$, $\ket{gr_-}$ and $\ket{er_-}$
are decoupled from the laser and microwave fields.
(c) Potential curves $E_l$, $E_m$, $E_u$ for the two-atom
microwave-dressed Rydberg states of Rb atoms with
$\ket{e} \equiv \ket{60S_{1/2},m_j = +\frac{1}{2}}$,
$\ket{r} \equiv \ket{60P_{3/2},m_j = +\frac{3}{2}}$, 
$\theta = \frac{\pi}{2}$, $\Om = 2\pi \times 0.1\:$GHz and $\De = -5 \Om$.
Dashed curves correspond to bare ($\Om \to 0$) energy levels
$E_{ee}$, $E_{er_\pm}$, $E_{rr}$ crossing at $R_{1,2,3}$.}
\label{fig:alsPots}
\end{figure}

Consider a pair of atoms with the Rydberg states $\ket{e}$
and $\ket{r}$ coupled by a microwave field with Rabi frequency $\Om$
and detuning $\De$, see Fig.~\ref{fig:alsPots}(a).
In the frame rotating with the microwave field frequency,
the interaction Hamiltonian for atom $j=1,2$ reads
$\mc{V}^j = -\hbar \Delta \sig_{rr}^j - \hbar \Om(\sig_{re}^j + \sig_{er}^j)$,
where $\sig_{\alpha \beta}^j \equiv \ket{\alpha_j}\bra{\beta_j}$ denote
the atomic operators. Atoms in states $\ket{e}$ and $\ket{r}$
interact via the vdW potentials
$\mc{W}_{e e} = \hbar \frac{C_6^{ee}}{R^6} \sig_{ee}^1 \otimes \sig_{ee}^2$
and $\mc{W}_{r r} = \hbar\frac{C_6^{rr}}{R^6} \sig_{rr}^1 \otimes \sig_{rr}^2$,
where $R$ is the interatomic distance and $C_6^{ee}$ and $C_6^{rr}$
are the corresponding vdW coefficients.
There is also a resonant DD (exchange) interaction between the atoms 
$\mc{D}_{er} = \hbar \sum_{i \neq j} \frac{C_3^{er}}{R^3} 
\sig_{re}^i \otimes \sig_{er}^j$, and an effective vdW interaction
$\mc{W}_{er} = \hbar \sum_{i \neq j} \frac{C_6^{er}}{R^6} 
\sig_{rr}^i \otimes  \sig_{ee}^j$ which arises from nonresonant DD interaction 
with shifted Rydberg level(s) \cite{supmat}. 
The total Hamiltonian for the pair of Rydberg atoms is 
$\mc{H}_{\mathrm{2Ry}} = \mc{V}^1 + \mc{V}^2 + \mc{W}_{e e} + \mc{W}_{r r}
+ \mc{D}_{er} + \mc{W}_{er}$.
Below we focus on the case of large detuning $|\De| \gg \Om$,
which exhibits rich structure and deep potential wells;
the (near-)resonant case $|\De| \lesssim \Om$ is discussed in \cite{supmat}.

Our two-atom basis set consists of states $\ket{ee} \equiv \ket{e_1 e_2}$,
$\ket{er_{\pm}} \equiv \frac{1}{\sqrt{2}} (\ket{e_1r_2} \pm \ket{r_1 e_2})$
and $\ket{rr} \equiv \ket{r_1 r_2}$ [Fig.~\ref{fig:alsPots}(b)].
When $\Om \to 0$, the (bare) energies of states $\ket{ee}$ and $\ket{rr}$ 
are given by $E_{ee}/\hbar = C_6^{ee}/R^6$ and 
$E_{rr}/\hbar = -2 \De + C_6^{rr}/R^6$, with $C_6^{ee} >0$
(repulsive vdW interaction) and $C_6^{rr} < 0$ (attractive vdW interaction)
while $\De < 0$ (red detuning). In turn, the resonant DD interaction
lifts the degeneracy of states $\ket{er_{\pm}}$ whose energies are
$E_{er_{\pm}}/\hbar = -\De \pm C_3^{er}/R^3 + C_6^{er}/R^6$, where 
$C_6^{er} >0$ is due to the effective vdW interaction \cite{supmat}.
Figure~\ref{fig:alsPots}(c) shows the dependence of bare energy levels 
$E_{\alpha \beta}$ on $R$. For vanishing vdW and DD interactions,
$R \to \infty$, $E_{ee} < E_{er_{\pm}} < E_{rr}$, while in the opposite limit
 of very strong (compared to $|\De|$) interactions, $R \to 0$, 
$E_{ee} > E_{er_+} > E_{er_-} > E_{rr}$ (in the rotating frame). 
There are three level crossing points of interest:
$E_{ee} = E_{er_+} \equiv E_{c1}$ at $R_1$,
$E_{ee} = E_{rr}  \equiv E_{c2} $ at $R_2$,  and
$E_{rr} = E_{er_+} \equiv E_{c3}$ at $R_3$, see \cite{supmat}.

Consider first the antisymmetric state $\ket{er_{-}}$. At large
distance, the potential $E_{er_-} \propto - R^{-3}$ is attractive 
due to the long-range resonant DD interaction, but at smaller distances,
the repulsive vdW interaction dominates, $E_{er_-} \propto R^{-6}$.
Hence, $E_{er_-}$ has a potential well, $\partial_R E_{er_-} = 0$,
around $R_- = \sqrt[3]{2C_6^{er}/C_3^{er}}$ where the vdW repulsion
overcomes the DD attraction [Fig.~\ref{fig:alsPots}(c)].
This simplified treatment captures the essential physics of the 
binding potential for DD interacting atoms presented in \cite{Kiffner2012}.
The antisymmetric state $\ket{er_{-}}$ will not play a role 
in our subsequent analysis since, for not too large dephasing,
it is decoupled from the other two-atom states and 
the microwave field, even when $\Om \neq 0$. 

We are thus left with three basis states $\ket{ee}$, $\ket{er_{+}}$
and $\ket{rr}$ coupled sequentially by the microwave field
with rate $\sqrt{2} \Om$, see Fig.~\ref{fig:alsPots}(b).
At large distances $R > R_{2,3}$, the vdW (and DD) interactions
are much weaker than $|\De|$ and the red-detuned ($\De < 0$) 
microwave field induces ac Stark shifts $\pm \frac{2 |\Om|^2}{\De}$ 
of levels $\ket{ee}$ and $\ket{rr}$ [Fig.~\ref{fig:alsPots}(c)]. 
At small distances, $R < R_{1,2}$, the vdW and DD shifts are so large that 
levels $\ket{ee}$, $\ket{rr}$ and $\ket{er_{+}}$ decouple from the field.
At the bare energy level crossing points $R_{1}$ and $R_3$,
the microwave field becomes resonant with the transitions
$\ket{er_{+}} \lra \ket{ee}$ and $\ket{er_{+}} \lra \ket{rr}$.
This leads to avoided crossings of the microwave-dressed energy levels
$E_u$ and $E_m$ which are repelled from $E_{c1,c3}$ by $\pm \sqrt{2} \Om$.
The upper potential curve $E_u$ has now a broad well near the crossing 
point $R_3$ of the bare levels $E_{er_+}$ and $E_{rr}$ determined by
the weakly repulsive DD interaction and strongly attractive vdW interaction. 
For the parameters used in Fig.~\ref{fig:alsPots}(c), at the bottom of the 
potential well the vibration frequency of the two-atom relative motion 
is $\nu_u \simeq 2 \pi \times 450\:$kHz \cite{supmat}.

Similarly, at the bare level crossing point $R_{2}$, the microwave
field couples states $\ket{ee} \lra \ket{rr}$, via nonresonant
intermediate state $\ket{er_{+}}$, with the two-photon Rabi frequency
$\Om^{(2)} = \hbar \frac{|\sqrt{2} \Om|^2}{E_{c2} - E_{er_{+}}(R_2)}$.
Now the microwave-dressed energy levels $E_l$ and $E_m$ are repelled
from $E_{c2}$ by $\pm \Om^{(2)}$, with the result that the middle
potential curve $E_m$ has a narrow well with a minimum near $E_{c2}$.
With the above parameters, $\Om^{(2)} = 2\pi \times 55\:$MHz and the 
two-atom relative vibrational frequency in the vicinity of the potential 
well minimum $R_m$ is $\nu_m \simeq 2 \pi \times 2 \:$MHz \cite{supmat}.
Note that the interatomic potentials of Fig.~\ref{fig:alsPots}(c) are 
azimuthally symmetric and robust with respect to small variations of 
the angle $\theta$ between the quantization axis and the two-atom 
separation vector, as shown in \cite{supmat}.

We next consider the excitation of the Rydberg states of atoms 
from the ground state $\ket{g}$ by a laser field acting on 
transition $\ket{g} \to \ket{e}$ with the Rabi frequency $\Om_p$ 
and detuning $\De_p$, see Fig.~\ref{fig:alsPots}(a).
The total Hamiltonian for the pair of atoms is now
$\mc{H} = \mc{H}_{\mathrm{2Ry}} + \mc{V}_p^1 + \mc{V}_p^2$
with the interaction Hamiltonian
$\mc{V}_p^j = \hbar \De_p \sig_{gg}^j -\hbar \Om_p (\sig_{eg}^j +\sig_{ge}^j)$.
We simulate the dynamics of the system using
the master equation for its density operator $\hrho$,
$\partial_t \hrho = -\frac{i}{\hbar} [\mc{H}, \hrho] +  \mc{L} \hrho$,
where the Liouvillian $\mc{L} \hrho = \sum_{j=1,2}
(\mc{L}_{g}^j \hrho + \mc{L}_{e}^j \hrho + \mc{L}_{r}^j \hrho)$, with 
$\mc{L}_{\alpha}^j \hrho = \hlf [2 \hL_{\alpha}^j \hrho \hL_{\alpha}^{j\dagger}
- \{ \hL_{\alpha}^{j\dagger} \hL_{\alpha}^j , \hrho \}]$, 
accounts for the relaxation
processes affecting the atoms \cite{QCQI2000,PLDP2007}.
These include the slow population decay of Rydberg states
$\ket{e},\ket{r}$ with rates $\Ga_{e,r}$, and the usually
more rapid decay of atomic coherences $\sig_{eg}$ and $\sig_{rg}$
with the total rate $\ga = \ga_{g} + \hlf \Ga_{e,r}$ ($\ga_{g} \gg \Ga_{e,r}$)
which originates from the laser phase fluctuations, Doppler shifts 
due to thermal atomic motion, and intermediate state decay when 
$\ket{g} \to \ket{e}$ is a two-photon transition \cite{NatPRLSaffman,%
NatPRLGrangier,Beguin2013,Kuzmich1213,Viteau2011,Schauss2012}.
The corresponding Lindblad generators are 
$\hL_e^j = \sqrt{\Ga_e} \sig_{ge}^j$, $\hL_r^j = \sqrt{\Ga_r} \sig_{gr}^j$
and $\hL_g^j = \sqrt{\ga_g/2} (\sig_{gg}^j - \sig_{ee}^j - \sig_{rr}^j)$.

\begin{figure}[t]
\centerline{\includegraphics[width=8.7cm]{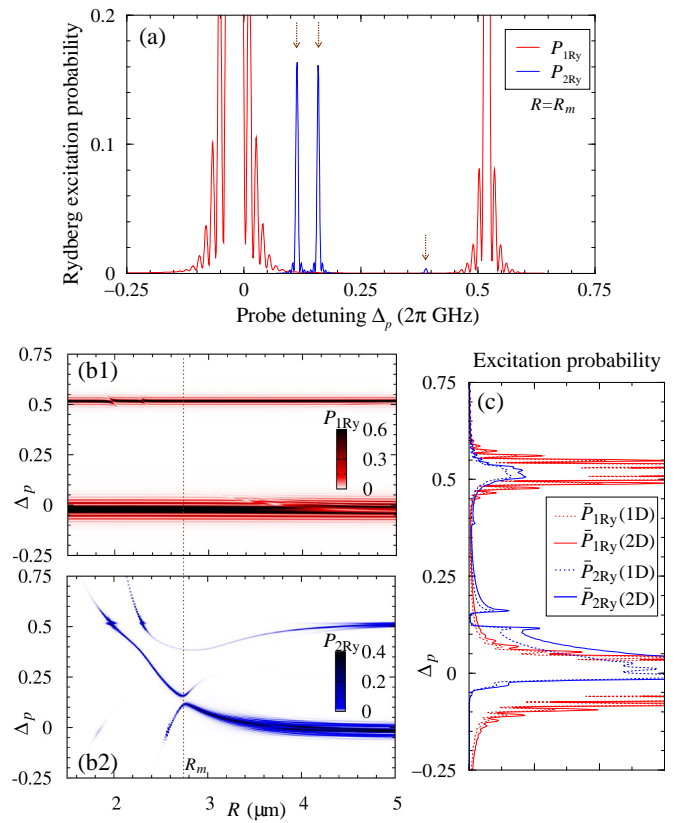}}
\caption{
Rydberg excitation spectra for two atoms excited from
the ground state $\ket{gg}$ by a smooth probe pulse $\Om_p(t)$.
(a)~One- and two-atom excitation probabilities $P_{\mathrm{1Ry}}$
and $P_{\mathrm{2Ry}}$ at interatomic distance $R_m = 2.74\:\mu$m.
(b1), (b2)~Density plots of $P_{\mathrm{1Ry}}, P_{\mathrm{2Ry}}$ vs distance $R$ 
[vertical dotted line marks $R=R_m$, cf. (a)].
(c) Spatially averaged excitation probabilities $\bar{P}_{\mathrm{1Ry}}$,
$\bar{P}_{\mathrm{2Ry}}$ for pairs of atoms in a 1D or 2D volume
of linear dimension $L = 5\:\mu$m.
In the simulations, the decay rates are $\Ga_{e,r} = 5\:$kHz
and $\ga_g = 2\pi\times 100\:$kHz, the probe Rabi frequency
$\Om_p = 2\pi \times 10\:$MHz ($0.1 \Om$) and duration
$\tau_p = 80\:$ns (flat-top pulse with Gaussian leading
and trailing edges of $10\:$ns duration);
other parameters are as in Fig.~\ref{fig:alsPots}(c).}
\label{fig:TyExPr}
\end{figure}

We assume that a short probe laser pulse $\Om_p(t)$ irradiates
the pair of free (thermal) atoms, followed by the detection of atoms 
in the Rydberg states, e.g., through dc field ionization. 
We vary the probe field frequency $\De_p$ and interatomic distance $R$
from pulse to pulse to obtain the Rydberg excitation probabilities
shown in Fig.~\ref{fig:TyExPr}. The probabilities of a 
single $P_{\mathrm{1Ry}} = \mathrm{tr}(\hrho \, \hat{\Pi}_{\mathrm{1Ry}})$ and
double $P_{\mathrm{2Ry}} = \mathrm{tr}(\hrho \, \hat{\Pi}_{\mathrm{2Ry}})$
excitations are defined through the projectors
$\hat{\Pi}_{\mathrm{1Ry}} \equiv \sum_{i \neq j} (\sig_{ee}^i + \sig_{rr}^i) 
\otimes \sig_{gg}^j$ and $\hat{\Pi}_{\mathrm{2Ry}} \equiv
(\sig_{ee}^1 + \sig_{rr}^1) \otimes (\sig_{ee}^2 + \sig_{rr}^2)$, 
where $\sig_{gg}^j + \sig_{ee}^j + \sig_{rr}^j = \mathds{1}_j$;
an actual experiment may or may not resolve the Rydberg state
and atom (or ion) number, therefore both $P_{\mathrm{1Ry}}$ and
$P_{\mathrm{2Ry}}$ are treated on equal footing.

Clearly, single atom excitation $P_{\mathrm{1Ry}}$, or pair excitation
$P_{\mathrm{2Ry}}$ at large interatomic distances $R$, are unaffected
by the Rydberg-state interactions. Scanning the frequency of the
laser field, we then probe the microwave field induced Autler-Townes
doublet $\la_{\pm} = -\hlf \De \pm \sqrt{\frac{1}{4}\De^2 + \Om^2}$ 
of the Rydberg states $\ket{e}$ and $\ket{r}$. 
For large microwave detuning $|\De| \gg \Om$, the probe field 
resonances are at $\De_p^{(-)} = \la_{-} \simeq - \frac{\Om^2}{|\De|}$ 
and $\De_p^{(+)} = \la_{+} \simeq |\De| + \frac{\Om^2}{|\De|}$, 
as expected [cf. Fig.~\ref{fig:alsPots}(c)].
Importantly, in the frequency region between $\De_p^{(-)}$ and $\De_p^{(+)}$,
destructive quantum interference leads to a dark resonance (EIT window) 
for the probe pulse \cite{stirap,EIT}:
If the pulse envelope varies adiabatically,
$\partial_t \Om_p < \Om_p |\la_{+} -\la_{-}|$, it does not
excite the bright eigenstates of the three-level atom,
which correspond to the Autler-Townes doublet when $\Om_p \ll \Om$.
During the interaction with the pulse, the populations of the Rydberg 
states are $\expv{\sig_{ee}} \simeq 0$ and 
$\expv{\sig_{rr}} \simeq \frac{\Om_p^2}{\Om^2}$,
but after the interaction, $\Om_p \to 0$, the atom returns to the
ground state with ideally unit probability $\expv{\sig_{gg}} \simeq 1$.
Atomic coherence relaxation $\ga$, however, reduces the transparency
causing residual population of the excited states.

Hence, within the EIT window for a single (non-interacting) atom,
we can probe the two-atom Rydberg resonances $E_{l,m,u}$.
In Fig.~\ref{fig:TyExPr}(a) we show the Rydberg excitation probabilities
versus probe detuning $\De_p$ for the atoms at distance $R=R_m$
of the $E_m$ potential minimum. Figure~\ref{fig:TyExPr}(b) summarizes
the results of our simulations for all (relevant) probe frequencies and
interatomic distances. Note that selective excitation of two-atom
Rydberg resonances from the ground state, $2\De_p = E_{l,m,u}/\hbar$,
requires two probe photons and such processes are therefore second order
in $\Om_p$, while single-atom resonances $\De_p = \la_{\pm}$ involve only 
one probe photon and are linear in $\Om_p$. An effective probe Rabi 
frequency for two-atom excitation is then much smaller than for 
a single-atom excitation, and in Fig.~\ref{fig:TyExPr} the smaller peaks 
of $P_{\mathrm{2Ry}}$ arise from a fractional ($\frac{1}{10}$) two-atom Rabi 
cycle, while the same amplitude and duration of the probe pulse results 
in several single-atom Rabi cycles leading to large and broad 
($\mathrm{sinc}$-shaped) peaks of $P_{\mathrm{1Ry}}$.

Note that a small excitation probability of the antisymmetric potential 
curve $E_{er_-}$ seen in Fig.~\ref{fig:TyExPr}(b2) (lower left corner) 
is due to dephasing $\gamma$ of individual atomic coherences.

If two atoms are confined in a uniform 1D (line) or 2D (disc) volume 
of linear dimension $L$, we can average the Rydberg excitation 
probabilities over the interatomic distances, 
$\bar{P}_{\mathrm{1,2Ry}} \equiv \int_0^L P_{\mathrm{1,2Ry}} \varrho(R) dR$,
where the corresponding probability densities for distances $R$
are given by $\varrho_{\mathrm{1D}}(R) = \frac{2(L-R)}{L^2}$ and
$\varrho_{\mathrm{2D}}(R) = \frac{8R}{\pi L^2} \left[ 2 \arccos \frac{R}{L}
- \frac{R}{2L} \sqrt{1 - \frac{R^2}{L^2} } \right]$.
As seen in Fig.~\ref{fig:TyExPr}(c), even after averaging over a large 
volume, we can still discern the structure of the two-atom Rydberg 
excitation probabilities $\bar{P}_{\mathrm{2Ry}}$ exhibiting an 
energy gap $2 \Om^{(2)} \simeq 2 \pi \times 0.1\:$GHz between
$\max{E_l}/2\hbar < \De_p < \min{E_m}/2\hbar$, not masked by
the single-atom excitation probability $\bar{P}_{\mathrm{1Ry}}$
within the EIT window.
Such features may still persist in low-density many-atom experiments,
provided the probability of having three- or more atoms within a few 
$\mu$m distance is small compared to the two-atom probability.

We now describe a potential application of the coherent, selective 
excitation of the Rydberg-dimer state for quantum information processing. 
Assume that a pair of cold atoms $1$ and $2$ are trapped at a relative
distance $R_0 \simeq R_m$ in an optical lattice \cite{Viteau2011,Schauss2012} 
or by far-detuned focused laser beams 
\cite{NatPRLSaffman,NatPRLGrangier,Beguin2013}. In each atom, 
long-lived states $\{\ket{s}, \ket{g}\}$ represent the qubit basis states.
The probe field resonantly couples the two-atom ground state $\ket{g_1 g_2}$
to the bound Rydberg-dimer state with the effective two-photon Rabi frequency 
$\Om_p^{(2)} \sim f \Om_p^2/\De_p$, where $f$ is the Franck-Condon overlap 
between the corresponding relative-coordinate wavefunctions \cite{supmat}. 
Atoms in state $\ket{s}$ are decoupled from the field, while single-atom 
Rydberg excitations from $\ket{g}$ are suppressed by the EIT mechanism. 
A pulse of effective area $\theta_p = 2\int_0^{\tau} \Om_p^{(2)}(t) dt = 2\pi$
thus leads to precisely one Rabi cycle between $\ket{g_1 g_2}$ and the 
Rydberg-dimer state, while all other initial states remain unaltered. 
The resulting $\pi$ phase shift of $\ket{g_1 g_2}$ corresponds to the 
two-qubit \textsc{cphase} logic gate \cite{QCQI2000,PLDP2007}.

\begin{figure}[t]
\centerline{\includegraphics[width=7.5cm]{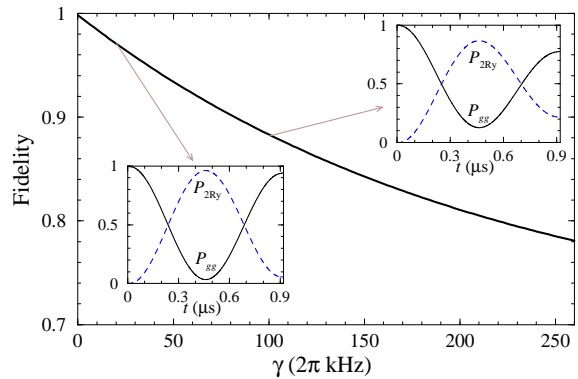}}
\caption{
Fidelity $F$ of \textsc{cphase} gate performed on atomic qubits
trapped at relative distance $R_m$ vs dephasing rate $\ga$.
Insets show probabilities $P_{gg}$ and $P_{\textrm{2Ry}}$ of two-atom
ground and bound Rydberg-dimer states during one Rabi cycle,
for $\gamma/(2\pi) = 20$ and $100\:$kHz.
Probe pulse detuning is $\De_p = 2\pi \times 159.3\:$MHz,
duration $\tau = 0.9\:\mu$s and its single-atom Rabi frequency
$\sqrt{f} \Om_p$ is scaled by the Franck-Condon factor $f=0.65$ \cite{supmat}.
Other parameters are as in Fig.~\ref{fig:TyExPr}(a).}
\label{fig:Fcphase}
\end{figure}

We have performed simulations of the two-atom gate using realistic 
experimental parameters \cite{supmat}. 
As seen in Fig.~\ref{fig:Fcphase}(insets),
sizable dephasing $\gamma$ detrimentally affects the amplitude of Rabi
oscillations between the two-atom internal-motional ground state and the 
lowest bound Rydberg-dimer state, and causes leakage of population to the 
vibrationally excited Rydberg-dimer states \cite{supmat}, reducing thereby 
the final population $P_{gg} = \expv{\sig_{gg}^1 \otimes \sig_{gg}^2}$ 
of state $\ket{g_1 g_2}$. To quantify the performance of the gate, 
we apply it to the input state $\ket{\Psi_{\textrm{in}}} 
= \hlf (\ket{s_1} +\ket{g_1}) \otimes (\ket{s_2} +\ket{g_2})$
containing equally weighted superposition of all two-qubit states. 
Ideally, the output state should be $\ket{\Psi_{\textrm{out}}} 
= \hlf (\ket{s_1 s_2} + \ket{s_1 g_2} + \ket{g_1 s_2} -  \ket{g_1 g_2})$.
In Fig.~\ref{fig:Fcphase} we show the resulting gate fidelity
$F = \bra{\Psi_{\textrm{out}}} \hrho(\tau) \ket{\Psi_{\textrm{out}}}$
which is close to unity for small $\gamma \ll \Om_p^{(2)}$, but
decreases with increasing $\gamma$, as expected.

Our quantum logic gate implementation complements previous 
proposals \cite{rydQIrev} in several ways. For moderate interatomic 
separation of a few micrometers, the strong interactions are typically
used for the blockade gate \cite{Jaksch2000} involving resonant excitation 
of only one Rydberg atom \cite{NatPRLSaffman,NatPRLGrangier,Beguin2013},
while at larger separation, both atoms can be excited and the interaction 
causes phase shift accumulated over time \cite{Jaksch2000,Durga2014}.
One can attempt to excite resonantly an anti-blockaded pair of strongly
interacting atoms while suppressing single-atom excitation by large detuning
(equal to half the interaction energy), but then the gradient of the vdW
potential amounts to strong mechanical force between the atoms, causing
motional decoherence and even excitation suppression \cite{WLi2013}.
In our implementation of a fast interaction gate, single atom excitations
are not merely suppressed by large detuning, but are almost completely
canceled by destructive quantum interference, and the gate is much less
vulnerable to motional decoherence since we resonantly excite the 
two-atom Rydberg state at a potential minimum. 

\begin{acknowledgments}
We are grateful to I. Lesanovsky and W. Li for valuable input and discussions,
and we acknowledge support from the Aarhus University Research Foundation
(AUFF), the FET-Open grant MALICIA (265522) and the Villum Foundation.
\end{acknowledgments}

\newpage

\section{Supplemental Material}

In these notes, we present details on the Rydberg states of the pair of 
atoms and their interactions, calculations of interatomic potential wells 
for large detuning of the microwave field as used in the main text,
interatomic potentials for different geometries and microwave field 
detunings, and the Franck-Condon factors for coherent excitation 
of trapped ground state atoms to the bound Rydberg dimer states.

\subsection{Atomic parameters and interatomic interactions}

\begin{figure}[b]
\centerline{\includegraphics[width=8.7cm]{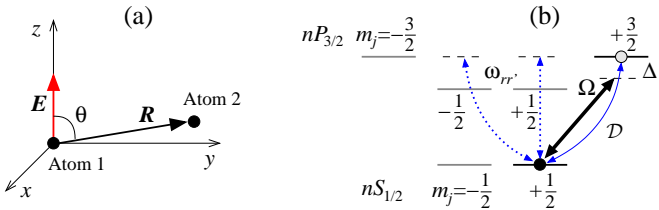}}
\caption{
(a) Geometry of the system of two Rydberg atoms:
$\theta$ is the angle between the relative position vector $\bs{R}$ and
the quantization axis $z$ defined by a static external field $\bs{E}$.
(b)~Level scheme of the magnetic sublevels $m_j$ of states $nS_{1/2}$
and $nP_{3/2}$ of each atom. The microwave field drives the transition
between $\ket{e} \equiv \ket{nS_{1/2},m_j=\frac{1}{2}}$ and
$\ket{r} \equiv \ket{nP_{3/2},m_j=\frac{3}{2}}$ with the Rabi frequency
$\Om$ and detuning $\De$. Resonant DD (exchange) interaction $\mc{D}$
can swap the states $\ket{e_1}\ket{r_2} \leftrightarrow \ket{r_1}\ket{e_2}$
of the two atoms, while the transitions to
$\ket{r'_{\pm}} \equiv \ket{nP_{3/2},m_j=\pm\frac{1}{2}}$
are non-resonant due to the $\bs{E}$ field induced differential
Stark shift $\om_{rr'}$.}
\label{fig:expgeom}
\end{figure}

The vector $\bs{R}$ connecting the positions of atoms 1 and 2
forms an angle $\theta$ with the quantization direction $\hat{z}$
defined by a static electric field $\bs{E}$, see Fig.~\ref{fig:expgeom}(a).
The interaction between the atomic dipoles $\bs{\wp}_1$ and $\bs{\wp}_2$
is given by
\begin{eqnarray*}
\mc{D} &=& \frac{1}{4 \pi \eps_0} \left[
\frac{\bs{\wp}_1 \cdot \bs{\wp}_2}{R^3}
- 3 \frac{(\bs{\wp}_1 \cdot \bs{R})(\bs{\wp}_2 \cdot \bs{R} ) }{R^5} \right] \\
&=& \frac{1}{4 \pi \eps_0 R^3} \Big[
\wp_{1+} \wp_{2-} + \wp_{1-} \wp_{2+} + \wp_{1z} \wp_{2z} (1-3 \cos^2 \theta)  \\
& &-\frac{3}{2} \sin^2 \theta (\wp_{1+} \wp_{2+} + \wp_{1+} \wp_{2-}
+ \wp_{1-} \wp_{2+} + \wp_{1-} \wp_{2-}) \\
& & \!\! -\frac{3}{\sqrt{2}} \sin \theta \cos \theta
(\wp_{1+} \wp_{2z} + \wp_{1-} \wp_{2z}
+ \wp_{1z} \wp_{2+} + \wp_{1z} \wp_{2-}) \Big] ,
\end{eqnarray*}
where $R \equiv |\bs{R}|$ while $\wp_{\pm, 0}$ denotes the dipole matrix element
for the atomic transition changing the magnetic quantum number $m_j$ (projection
of the total angular momentum $J$ onto $\hat{z}$) by $\De m_j = \pm 1, 0$.
The interaction is invariant under rotation about the $\hat{z}$ axis.

We assume alkali atoms. The two Rydberg states of each atom
are represented by $\ket{e} \equiv \ket{nS_{1/2},m_j = +\frac{1}{2}}$
and $\ket{r} \equiv \ket{nP_{3/2},m_j = +\frac{3}{2}}$ with the principal
quantum number $n$ ($\gtrsim 40$), see Fig.~\ref{fig:expgeom}(b).
A $\sigma_+$-polarized microwave field $\mc{E}_{\mathrm{mw}}$ of frequency
$\omega_{\mathrm{mw}}$ drives the transition $\ket{e} \lra \ket{r}$ with
the Rabi frequency $\Om = \wp_{er} \mc{E}_{\mathrm{mw}}/\hbar$ and
detuning $\De = \omega_{\mathrm{mw}} - \omega_{re}$.
We can recast the DD interaction as 
\begin{eqnarray}
\mc{D} &=& \frac{2 - 3\sin^2 \theta}{8 \pi \eps_0 R^3}
(\wp_{1+} \wp_{2-} + \wp_{1-} \wp_{2+}) 
\nonumber \\  & &
- \frac{ 3 \sin^2 \theta}{8 \pi \eps_0 R^3}
\Big[ \wp_{1+} \wp_{2+} + \wp_{1-} \wp_{2-} 
+ \frac{\sqrt{2} \cos \theta}{\sin \theta}   
\nonumber \\ & & \qquad 
\times (\wp_{1+} \wp_{2z} + \wp_{1-} \wp_{2z}
+ \wp_{1z} \wp_{2+} + \wp_{1z} \wp_{2-}) \Big] , \quad
\label{eqsup:resnonresDD}
\end{eqnarray}
where we neglected the term $\wp_{1z} \wp_{2z}$ since states 
$\ket{r'_{\pm}} \equiv \ket{nP_{3/2},m_j= \pm \frac{1}{2}}$ 
are not populated by the microwave field.
The first term on the right-hand side of Eq.~(\ref{eqsup:resnonresDD})
describes the resonant DD exchange interaction 
$\ket{e_{1(2)}r_{2(1)}} \lra \ket{r_{1(2)}e_{2(1)}}$ denoted 
by $\mc{D}_{eg}$ in the main text. The transition matrix elements
are $\wp_{\pm} = \mp \frac{1}{\sqrt{3}} (nS_{1/2} || \wp || nP_{3/2})$,
where the numerical prefactor corresponds the angular part
while the reduced matrix element in the semiclassical
approximation \cite{Kaulakys1995_s} is given by
$(nS_{1/2} || \wp || nP_{3/2}) \approx -\frac{3}{2} n^{*2}$
with $n^* = n- \de_S $ the effective principal quantum number.
The DD coefficient is then 
$C_3^{er} \approx \frac{3 (3 \sin^2 \theta -2)}{32 \pi \eps_0} n^{*4}$.
The second term on the right-hand side of Eq.~(\ref{eqsup:resnonresDD})
corresponds to the DD interaction coupling states
$\ket{r_{1(2)}e_{2(1)}}$ to $\ket{e_{1(2)}r'_{2(1)}}$ with rates
$C_3^{re,er'_{-}} \approx \frac{3 \sqrt{3} \sin^2 \theta}{32 \pi \eps_0} n^{*4}$
and $C_3^{re,er'_{+}} \approx \frac{3 \sqrt{3} \sin \theta \cos \theta }
{16 \pi \eps_0} n^{*4}$. This process could populate states $\ket{r'}$ 
outside the two-level subspace $\{ \ket{e},\ket{r} \}$, but we assume that 
this leakage is suppressed by the external electric $\bs{E}$ (or magnetic) 
field inducing differential Stark (or Zeeman) shift $\om_{rr'}$ 
between levels $\ket{r}$ and $\ket{r'_{\pm}}$, see Fig.~\ref{fig:expgeom}(b).
The non-resonant DD interaction induces, however, a second-order
level shift of states $\ket{r_{1(2)}e_{2(1)}}$, which we account for
as an effective vdW interaction $\mc{W}_{er} = \hbar \frac{C_6^{er}}{R^6}
\sig_{rr}^{1(2)} \otimes \sig_{ee}^{2(1)}$ with the coefficient
$C_6^{er} = \sum_{r'} \frac{|C_3^{re,er'}|^2}{\om_{rr'}}$, which is positive 
(repulsive vdW interaction) if levels $\ket{r'}$ are lower than 
$\ket{r}$. We note that this is a rather simplistic approximation which,
strictly speaking, is valid only at large enough distances
$R > \sqrt[3]{\frac{|C_3^{re,er'}|}{\om_{rr'}}}$.
The combination of resonant and nonresonant DD interactions
and the resulting binding potential was analyzed rigorously in
\cite{Kiffner2012_s}, while our simplification facilitates derivation
of the hitherto unexplored binding potentials originating from microwave
dressing of vdW interacting states $\ket{e_1 e_2}$ and $\ket{r_1 r_2}$.

To be specific, we take Rb atoms in $n=60$ states. The quantum
defects for the $S_{1/2}$ and $P_{3/2}$ states are $\de_S = 3.13109$ and
$\de_P = 2.65145$ \cite{RydAtoms_s}, with which the (unshifted)
$\ket{e} \to \ket{r}$ transition frequency is
$\om_{re} \simeq 2 \pi \times 17\:$GHz. In the main text, 
we consider the case $\theta = \frac{\pi}{2}$ when the 
quantization axis $\hat{z}$ is perpendicular to the two-atom 
separation vector $\bs{R}$. For the DD interaction, we then 
obtain $C_3^{er} \approx 2 \pi \times 3.8\:$GHz$\:\mu$m$^3$
and assume $C_6^{er} = 2 \pi \times 3\:$GHz$\:\mu$m$^6$, while
the coefficients for the vdW potentials $\mc{W}_{ee}$ and $\mc{W}_{rr}$
are $C_6^{ee} \simeq 2 \pi \times 140\:$GHz$\:\mu$m$^6$ (repulsion) and
$C_6^{rr} \simeq - 2 \pi \times 295\:$GHz$\:\mu$m$^6$ (attraction)
\cite{Reinhard2007_s,Singer2005_s}. For other values of $\theta$ (see below),
the angular dependence of the DD and vdW coefficients is 
$C_3^{er}(\theta) \propto (3 \sin^2 \theta -2)$ and 
$C_6^{rr}(\theta) \propto \sin^2 \theta$,
while $C_6^{ee}$ remains isotropic \cite{Reinhard2007_s}.

\subsection{Crossing points of $E_{\alpha \beta}$ and potential minima of $E_{m,u}$}

For vanishing microwave field amplitude, $\Om \to 0$, the crossing points
of the bare two-atom potentials $E_{ee}$, $E_{er_+}$, and $E_{rr}$ are
\begin{eqnarray*}
& E_{ee} = E_{er_+} \equiv
E_{c1} = \frac{4 \hbar \De^2 C_6^{ee}}
{\left[\sqrt{(C_3^{er})^2 + 4 |\De| (C_6^{ee} - C_6^{er}) } - C_3^{er} \right]^2} \\
& \mathrm{at} \;\;
R_1 = \left[ \sqrt{ \left( \frac{C_3^{er}}{2\De} \right)^2
+ \frac{C_6^{ee} - C_6^{er}}{|\De|} } - \frac{C_3^{er}}{2|\De|} \right]^{1/3} ,
\end{eqnarray*}
\begin{eqnarray*}
& E_{rr} = E_{er_+} \equiv
E_{c3} = -2 \hbar \De + \frac{4 \hbar \De^2 C_6^{rr}}
{\left[\sqrt{(C_3^{er})^2 + 4 |\De| (C_6^{er} - C_6^{rr})} + C_3^{er} \right]^2} \\
 & \mathrm{at} \; \;
R_3 = \left[ \sqrt{ \left( \frac{C_3^{er}}{2\De} \right)^2
+ \frac{C_6^{er} - C_6^{rr}}{|\De|}} + \frac{C_3^{er}}{2|\De|} \right]^{1/3} ,
\end{eqnarray*}
and
\begin{eqnarray*}
& E_{ee} = E_{rr} \equiv
E_{c2} = \frac{2 \hbar |\De| C_6^{ee}}{C_6^{ee} - C_6^{rr}}
\;\; \mathrm{at} \;\; R_2 = \sqrt[6]{ \frac{C_6^{ee} - C_6^{rr}}{2|\De|}} .
\end{eqnarray*}
With the atomic parameters listed above, we have
$R_1 \simeq 2.36\:\mu$m, $R_2 \simeq 2.75\:\mu$m, and
$R_3 \simeq 3.05\:\mu$m.

In Fig.~1(c) of the main text, the potential energy curves
$E_m$ and $E_u$ have potential minima. Consider first the potential
well on the $E_m$ curve which is defined by the bare energy levels
$E_{ee}$ and $E_{rr}$ crossing at $R_2$. The microwave field couples states
$\ket{ee}$ and $\ket{rr}$ by a two-photon transition via non-resonant
intermediate state $\ket{er_{+}}$ with the effective Rabi frequency
$\Om^{(2)} = \frac{\hbar |\sqrt{2} \Om|^2}{E_{c2} - E_{er_+}}$.
At $R_2$, we have $E_{er_+} \approx \hbar |\De| +
\frac{\hbar C_3^{er} \sqrt{2 |\De|}}{\sqrt{C_6^{ee} - C_6^{rr}} }$,
which upon substitution leads to
\[
\Om^{(2)} = \frac{2|\Om|^2 (C_6^{ee} - C_6^{rr})}
{|\De| (C_6^{ee} + C_6^{rr}) - C_3^{er} \sqrt{2\De(C_6^{ee} - C_6^{rr})}}.
\]
For our parameters, both terms in the denominator are comparable and hence
the DD interaction cannot be neglected. We obtain
$|\Om^{(2)}| \simeq 2\pi \times 55\:$MHz.

In the vicinity of $E_{c2}$ and $R_2$, we thus have an effective
two-level system described by Hamiltonian
\[
\mc{H}_{\mathrm{eff}} = \left[ \begin{array}{cc}
E^{(1)}_{ee} & \hbar \Om^{(2)}  \\
\hbar \Om^{(2)} & E^{(1)}_{rr}
\end{array} \right] ,
\]
where $E^{(1)}_{\alpha \alpha} = E_{c2} +  \eta_{\alpha \alpha } (R-R_2)$ are
linear approximations for the bare energies of the corresponding
states ($\alpha \alpha =ee,rr$), with $\eta_{\alpha \alpha }
= \left.\frac{\partial E_{\alpha \alpha}}{\partial R} \right|_{R_2}$.
The binding potential is $E_{m} = \hlf[E^{(1)}_{ee} + E^{(1)}_{rr}]
+ \sqrt{\frac{1}{4}[E^{(1)}_{ee} - E^{(1)}_{rr}]^2 +|\hbar \Om^{(2)}|^2}$,
whose minimum $R_m$ is found by $\partial_R E_{m} =0$.
Expanding $E_{m}$ up to the second order in $R$ around $R_m$, 
we obtain the harmonic potential 
$E_{m} \approx (E_{c2} + \hbar \Om^{(2)}) + \hlf \kappa_m (R-R_m)^2$ with
\[
\kappa_m = \frac{2}{\hbar |\Om^{(2)}|}
\frac{|\eta_{ee} \eta_{rr}|^{3/2}}{|\eta_{ee}|+ |\eta_{rr}|}
= \frac{2\hbar}{R_2^2} \frac{(12 \De)^{2} (C_6^{ee} |C_6^{rr}|)^{3/2}}
{|\Om^{(2)}| (C_6^{ee} + |C_6^{rr}|)^{3}} .
\]
Since $\kappa_m = \mu \nu_m^2$, with $\mu$ the reduced mass,
the relative vibration frequency of two identical atoms of mass
$M_{\mathrm{at}}$ is $\nu_m \approx \sqrt{2 \kappa_m/M_{\mathrm{at}}}$.
With the parameters for $^{87}$Rb atoms listed above, we have
$\nu_m \simeq 2 \pi \times 2 \:$MHz around $R_m = 2.74\:\mu$m.

Since the depth of the binding potential $\sim |\De|$ is much larger
than the vibrational frequency $\nu_m$, the harmonic approximation 
holds for many vibrational states $n \ll |\De|/\nu_m \sim 10^2$ of 
the Rydberg dimer, whose energies are given by 
$\veps_n \simeq \hbar \nu_m(\hlf+n)$ while the corresponding wavefunctions
are
\begin{equation}
\chi_m(R,n) = \frac{2^{-n/2}}{\sqrt{n!}}
\left(\frac{1}{\pi \Sigma_m^2} \right)^{\frac{1}{4}}
e^{-\frac{(R-R_m)^2}{2 \Sigma_m^2}}
H_n \left( \frac{R}{\sqrt{\pi} \Sigma_m} \right),
\end{equation}
where $\Sigma_m = \sqrt{2\hbar/M_{\mathrm{at}} \nu_m}$ and $H_n(R)$ are the 
Hermite polynomials. 

Consider now the $E_u$ potential energy curve.
The potential well is bounded by $E_{ee}$ on the left and $E_{rr}$
on the right sides, and $E_{er_+}$ at the bottom. Since with
increasing distance $R$ the DD interaction slowly lowers $E_{er_+}$
while $E_{rr}$ rapidly approaches $2|\De|$, the minimum of the potential
well is located above, and to the left from, the energy level crossing
point $E_{c3}$, $R_3$ of the bare states $\ket{er_+}$ and $\ket{rr}$
coupled by the microwave field with the Rabi frequency $\sqrt{2} \Om$.
Now the Hamiltonian for the effective two-level system is
\[
\mc{H}_{\mathrm{eff}} = \left[ \begin{array}{cc}
E^{(1)}_{er_+} & \hbar \sqrt{2} \Om  \\
\hbar \sqrt{2} \Om & E^{(1)}_{rr}
\end{array} \right] ,
\]
where $E^{(1)}_{\alpha \beta} = E_{c3} +  \eta_{\alpha \beta} (R-R_3)$ with
$\eta_{\alpha \beta}=\left.\frac{\partial E_{\alpha \beta}}{\partial R} \right|_{R_3}$.
Proceeding as above, we obtain for the potential well
$E_{u} \approx (E_{c3} + \hbar \sqrt{2} \Om) + \hlf \kappa_u (R-R_u)^2$ with
\[
\kappa_u = \frac{2}{\hbar \sqrt{2} |\Om|}
\frac{|\eta_{er_+} \eta_{rr}|^{3/2}}{|\eta_{er_+}|+ |\eta_{rr}|}
\simeq \frac{2\hbar}{R_3^2} \frac{9 |\De|^{5/4} (C_3^{er})^{3/2}}
{|\Om| |C_6^{rr}|^{3/4}} ,
\]
where we assumed $R_3 \simeq \sqrt[6]{|C_6^{rr}/\De|}$ and
neglected $C_3^{er} R_3^3$ in comparison with $2 |C_6^{rr}|$
since the DD interaction varies slowly compared to the vdW interaction.
For the vibration frequency of the two-atom relative motion
around $R_u = 2.85\:\mu$m, we then obtain
$\nu_u \approx 2 \pi \times 450\:$kHz.

\subsection{Two-atom potentials for $\theta \neq \frac{\pi}{2}$}

\begin{figure}[t]
\centerline{\includegraphics[width=8.7cm]{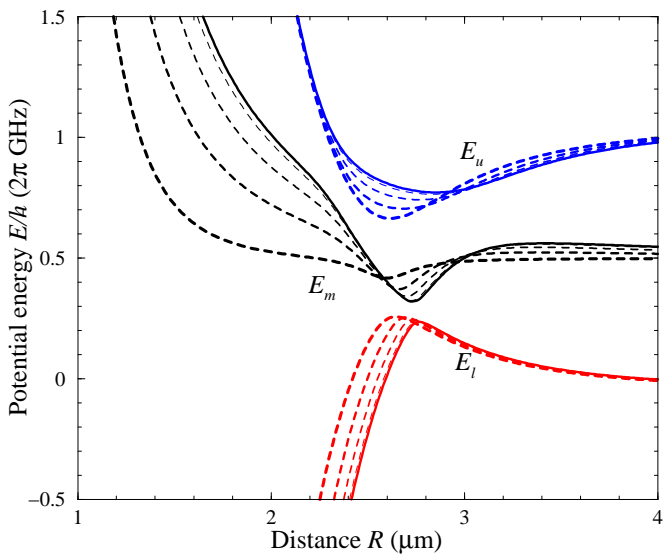}}
\caption{
Potential curves $E_{l,m,u}$ for the two-atom microwave-dressed Rydberg
states, for the same parameters as in the main text but different $\theta$:
Solid curves are for $\theta = \frac{\pi}{2}$ [as in Fig.~1(c)], with the 
increment of $\de \theta = 0.05 \pi$ for progressively thicker dashed curves,
up to $\theta = 0.7 \pi$.}
\label{fig:Potstheta}
\end{figure}

In the main text, we consider the case of $\theta = \frac{\pi}{2}$ 
corresponding to the quantization axis $\hat{z}$ being perpendicular to 
the two-atom separation vector $\bs{R}$ (cf. Fig.~\ref{fig:expgeom} above).
In Fig.~\ref{fig:Potstheta} we show the potential curves $E_{l,m,u}$
for values of $\theta \neq \frac{\pi}{2}$. As seen, small variations
of angle $\theta$ around $\frac{\pi}{2}$ change the interatomic potentials
only little. With increasing $|\theta - \frac{\pi}{2}|$, the broad potential
well on the upper curve $E_{u}$ becomes deeper and moves towards smaller
distances $R$. Simultaneously, the potential well on the middle curve $E_{m}$
gets shallower until it nearly disappears for 
$\theta \gtrsim 0.7 \pi$ when the strength of 
the attractive potential $C_6^{rr}(\theta) \propto \sin^2 \theta$
becomes comparable to, or smaller than, that of the repulsive 
potential $C_6^{ee}$, which does not depend on $\theta$.

The interatomic potentials are azimuthally symmetric, i.e., 
the potential curves $E_{l,m,u}$ are invariant under rotation of 
the interatomic separation vector $\bs{R}$ about the $\hat{z}$ axis,
which would draw 2D doughnut shaped potential surfaces.

\subsection{Two-atom potentials for near-resonant microwave field}

In the main text, we considered the case of a large negative detuning
$\De <-|\Om|$ of the microwave field from the single-atom transition 
resonance $\ket{e} \to \ket{r}$, which leads to two binding potentials 
with the depths $\sim |\De|$. Here we outline the (near-) resonant
case $|\De| \lesssim |\Om|$.

\begin{figure}[t]
\centerline{\includegraphics[width=8.7cm]{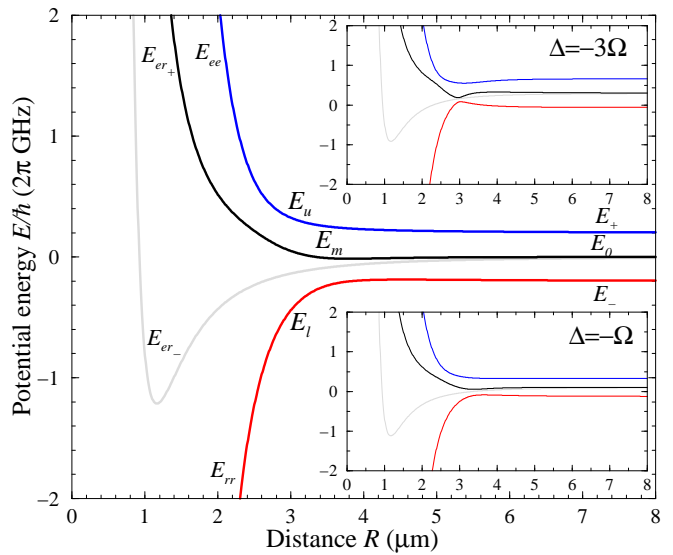}}
\caption{
Potential curves $E_{l,m,u}$ for the same parameters as in Fig.~1(c) 
of the main text, but with $\De = 0$ (main panel), and 
$\De = -\Om$ (lower inset) and $\De = -3\Om$ (upper inset).}
\label{fig:PotsD0}
\end{figure}

In Fig.~\ref{fig:PotsD0} (main panel) we show the potential energy
curves for $\De = 0$ and $\Om \neq 0$. Consider again the bare two-atom
states $\ket{ee}$, $\ket{er_{\pm}}$, and $\ket{rr}$. The antisymmetric
state $\ket{er_{-}}$ is decoupled from the field at any relative
distance $R$, so its behavior is the same for any $\De$ which simply
shifts the zero-point energy. At large distances
$R > R_b \simeq \sqrt[6]{C_6^{\alpha \beta}/\Om},\sqrt[3]{C_3^{er}/\Om}$,
all two-atom states $\ket{\alpha \beta}$ have the same energy
$E_{\alpha \beta} = 0$ (in the frame rotating with the microwave frequency).
The microwave field couples resonantly the transitions
$\ket{ee} \lra \ket{er_{+}}$ and $\ket{er_{+}} \lra \ket{rr}$ with
the same Rabi frequency $\sqrt{2} \Om$. The eigenstates of the system
are then $\ket{\Psi_0} = \frac{1}{\sqrt{2}} (\ket{ee} - \ket{rr})$
and  $\ket{\Psi_{\pm}} = \frac{1}{2} (\ket{ee} \pm \ket{er_{+}} + \ket{rr})$
with the corresponding energies $E_0 = 0$ and $E_{\pm}/\hbar = \pm 2 \Om$,
as can be observed in Fig.~\ref{fig:PotsD0}. At small distances
$R < R_b$, the vdW (and DD) shifted states $\ket{ee}$ and $\ket{rr}$
(and $\ket{er_{+}}$) are completely decoupled from the microwave field.
The transition between the two regimes occurs in the vicinity
of $R = R_b$. If the attractive vdW interaction is stronger than the
repulsive one, $|C_6^{rr}| > C_6^{ee}$, there is a shallow potential
well on the middle curve $E_m$ (in the opposite case of
$|C_6^{rr}| < C_6^{ee}$ there would be a small hump).

With lowering the frequency of the microwave field to increase
the absolute (but negative) value of the detuning $\De \sim -|\Om|$, 
the potential well on the $E_m$ curve becomes more pronounced 
[Fig.~\ref{fig:PotsD0} lower inset];
for still larger (negative) values of $\De \lesssim -\Om$,
the potential energy curves approach those described in the main text
[compare Fig.~\ref{fig:PotsD0} upper inset with Fig.~1(c) of the main text].

There are no potential wells for positive detuning $\De > 0$,
if $C_6^{rr} < 0$ and $C_6^{ee} > 0$. The situation would be reverse
for repulsive vdW interaction between the upper Rydberg states $C_6^{rr} > 0$
and attractive interaction between the lower states $C_6^{ee} < 0$,
i.e., we would need $\De > 0$ to obtain binding potentials.

\subsection{Excitation of trapped ground state atoms to the Rydberg-dimer state}

We assume that the atoms $j=1,2$ in the ground internal state $\ket{g}$
are localized around positions $r_{j,0}$ of two separate traps.
For cold atoms, we can approximate the spatial wavefunctions of the
atoms $\psi_{j}$ by the ground-state wavefunction of a harmonic oscillator
\[
\psi_{j}(r_{j}) \approx \left(\frac{1}{\pi \sigma^2} \right)^{\frac{1}{4}}
e^{-\frac{(r_j-r_{j,0})^2}{2 \sigma^2}} ,
\]
where the width $\sigma = \sqrt{\hbar/M_{\mathrm{at}} \nu}$ is expressed
through the vibrational frequency $\nu$ assumed to be the same for both atoms.
The two-atom wavefunction
$\Psi_{12} = \psi_{1}\psi_{2}$ can be expressed in terms of the center of mass
$\bar{r} = \hlf (r_{1} + r_{2})$ and relative $R = r_{2} - r_{1}$ coordinates
as $\Psi_{12} (\bar{r},R) =  \phi(\bar{r}) \, \chi(R)$ with
\begin{eqnarray}
\phi(\bar{r}) &=& \left(\frac{1}{\pi \bar{\sigma}^2} \right)^{\frac{1}{4}}
e^{-\frac{(\bar{r}- \bar{r}_{0})^2}{2 \bar{\sigma}^2}} , \\
\chi(R) &=& \left(\frac{1}{\pi \Sigma^2} \right)^{\frac{1}{4}}
e^{-\frac{(R-R_{0})^2}{2 \Sigma^2}}  ,
\end{eqnarray}
where $\bar{r}_0 = \hlf (r_{1,0} + r_{2,0})$ and $R_0 = r_{2,0} - r_{1,0}$,
while $\bar{\sigma} = \frac{1}{\sqrt{2}} \sigma$ and
$\Sigma = \sqrt{2}\sigma$.

Our aim is to coherently and reversibly excite the two ground state atoms 
to a single Rydberg-dimer state on the $E_m$ potential energy curve.
We therefore assume an appropriate distance between the trapped
atoms $R_0 \simeq R_m$ and apply the probe field $\Om_p$ which
is two-photon resonant between the internal-motional ground
state $\ket{G} = \ket{g_1 g_2} \otimes \chi(R)$ and the lowest 
Rydberg-dimer state $\ket{D_m} = \ket{\Psi_m} \otimes \chi_m(R)$ with
\begin{equation}
\chi_m(R) \equiv \chi_m(R,0) = \left(\frac{1}{\pi \Sigma_m^2} \right)^{\frac{1}{4}}
e^{-\frac{(R-R_m)^2}{2 \Sigma_m^2}} , \label{chi_m0}
\end{equation}
where $\Sigma_m = \sqrt{2\hbar/M_{\mathrm{at}} \nu_m}$.
The corresponding Franck–Condon factor for the transition
$\ket{G} \to \ket{D_m}$ is
\[
f = \int_0^{\infty} \!\! \chi^*(R) \chi_m(R) \, dR =
\left(\frac{2\Sigma \Sigma_m}{\Sigma^2 + \Sigma_m^2} \right)^{\frac{1}{2}}
e^{-\frac{(R_0 -R_m)^2}{2 (\Sigma^2 + \Sigma_m^2)} } ,
\]
where we included the contribution of a possible mismatch between
the equilibrium interatomic distance $R_0$ in the ground state traps
and the position $R_m$ of the two-atom potential minimum in the
Rydberg state. Taking the trap frequency $\nu \simeq 2\pi \times 100\:$kHz
for the ground-state $^{87}$Rb atoms, while for the Rydberg-dimer state
we have $\nu_m \simeq 2\pi \times 2\:$MHz (see above), we obtain
$\Sigma \simeq 48\:$nm and $\Sigma_m \simeq 10.7\:$nm.
Assuming small mismatch $|R_0 -R_m| \ll \Sigma_m$ leads to
the Franck-Condon factor $f \simeq 0.65$ used in Fig.~3 of the main text.

Under the harmonic approximation for the $E_m$ binding potential,
the Franck-Condon factors for the transitions between the internal-motional 
ground state $\ket{G}$ of a pair of atoms and $n$th vibrationally excited
state $\ket{D_m^{(n)}} = \ket{\Psi_m} \otimes \chi_m(R,n)$ of the Rydberg
dimer are given by $f(n) = \int_0^{\infty} \!\! \chi^*(R) \chi_m(R,n) \, dR$.
With the above parameters and $|R_0 -R_m| =0$, we obtain for the even excited 
states $f(2,4,6,8,10,\ldots) =\{0.417,0.327,0.27,0.229,0.197,\ldots\}$,
while for all the odd excited states the Franck-Condon factors vanish, 
$f(1,3,\ldots) = 0$. A mismatch between $R_0$ and $R_m$, however, can make 
the transitions to the odd-$n$ states allowed: e.g., for 
$|R_0 -R_m| \simeq 1\:$nm we obtain $f(1,3,\ldots) \simeq 0.045$. 

\begin{figure}[t]
\centerline{\includegraphics[width=8.7cm]{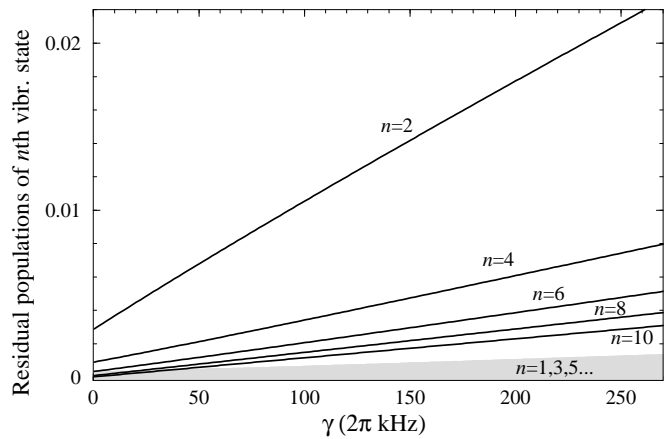}}
\caption{
Residual populations of the $n \geq 1$ vibrationally excited states
of the Rydberg dimer after the effective $2\pi$ pulse implementing 
the \textsc{cphase} gate of Fig.~3 of the main text. The solid lines 
show the populations of the even-$n$ states vs the dephasing rate $\gamma$,
while the shaded area indicates the maximal possible population of 
the odd-$n$ states for $|R_0-R_m| \lesssim 0.1 \Sigma_m$ mismatch
of the trapping distance of the atom pair.}
\label{fig:ResPop}
\end{figure}

When the probe field with the effective two-photon Rabi frequency 
$\Om_p^{(2)} \sim f(n) \Om_p^2/\De_p$ resonantly couples the pair 
of ground state atoms to the lowest Rydberg-dimer state ($n=0$), 
as in Fig.~3 of the main text, the vibrationally excited states $n \geq 1$ 
are detuned by $n \nu_m$, which, together with the smaller Franck-Condon 
factors $f(n)$, suppresses their excitation. In Fig.~\ref{fig:ResPop} above, 
we show the residual excitation probabilities of $n \geq 1$ vibrational states
at the end of the effective $\theta_p = 2\pi$ pulse implementing 
the \textsc{cphase} gate. Due the small linewidths of the two-photon 
transitions to the  Rydberg-dimer states, the populations of the even-$n$ 
states are below $1\%$ for $\gamma/(2\pi) < 100\:$kHz, while the excitation 
probability of the closest $n=1$ state due to a possible 
$|R_0 -R_m| \lesssim 1\:$nm uncertainty in trap distance of the ground state 
atoms is less than $6.5 \times 10^{-4}$; higher odd-$n$ state can acquire 
even less population. Hence, during the gate execution the leakage of 
population out of the qubit subspace into vibrationally excited Rydberg
dimer states is insignificant for moderate values of dephasing $\gamma$.  

We finally note that during the gate operation the center-of-mass and 
spatial orientation of the Rydberg dimer can disperse freely, resulting 
in motional decoherenece. An additional weak trapping potential for 
the Rydberg state atoms \cite{rydQIrev} can compensate this dispersion 
insuring complete return of the wavefunction of the two atoms to the 
trapped ground state.

\end{document}